\documentclass{mq}

\title[The Milliquas Catalogue]{The Million Quasars (Milliquas) Catalogue, v6.4}
\author[E. Flesch]{\textbf{Eric W. Flesch }$^{A,B}$\\
\\
\affil{$^A$PO Box 15, Dannevirke 4942, New Zealand}
\affil{$^B$Email: eric@flesch.org}}

\begin{document}

\begin{abstract}
Announcing the release v6.4 of the Milliquas (Million Quasars) quasar catalogue which presents all published quasars to 11 December 2019, including SDSS-DR16.  Its totals are 757\,991 type-I QSOs/AGN and approx 1.1M high-confidence (80\%+ likelihood) quasar candidates from SDSS-based \& AllWISE photometric quasar catalogs, plus all-sky radio/X-ray associated candidates available only here.  Type-II and Bl Lac objects are also included, plus candidates/galaxies with double radio lobes (so calculated), bringing the total count to 1\,968\,377.  \textit{Gaia}-DR2 astrometry is given for most objects.  The catalogue is available on its home page and on NASA HEASARC. 
\end{abstract}


\maketitle

\section{Introduction}

Milliquas is currently the only quasar catalogue to keep abreast of all the latest quasar discoveries from published papers large \& small; this edition v6.4 is current to 11 December 2019 and includes the SDSS-DR16 pipeline catalogue which adds $\approx$160\,000 quasars to the literature.  The criteria for accepting pipeline quasars and other specifications are as given in the Half Million Quasars catalog (HMQ: Flesch 2015) and references therein, but some modifications are listed below.  

Milliquas can be downloaded from its home page\footnote{http://quasars.org/milliquas.htm} or from NASA HEASARC\footnote{https://heasarc.gsfc.nasa.gov/W3Browse/all/milliquas.html} which also provides a query page.  Its ReadMe, also available there, gives essential information about the data.  If using Milliquas in published research, please cite to this arXiv document.

\section{Changes in Milliquas specifications since the HMQ}

The processing rules of the HMQ \shortcite{HMQ} are still used but have evolved somewhat due to better quality data in recent years, plus more rigorous use of SDSS pipeline subclasses, and handling anti-selection where SDSS/LAMOST pipeline data are not included into their own manually-vetted quasar catalogs.  

The optical background has been switched to that of the All-Sky Portable (ASP) optical catalog \cite{ASP} which is comprised of APM/USNO-B/SDSS sourced data to an astrometric precision of 0.1 arcsecond and a photometric precision of 0.01 magnitude, and also $\approx$63\% of the Milliquas astrometry is now from the very accurate \textit{Gaia}-DR2 \cite{GAIA2}.  All radio/X-ray associations have been recalculated onto the optical background of Figure 3 of Flesch \shortcite{ASP}.  Associations to faint SDSS objects are now calculated onto a usually-higher background density (because of binning by optical PSF \& red-blue colour), thus their likelihoods usually reduced and consequently $\approx$20K candidates fell below pQSO=80\% (the acceptance threshold for Milliquas) and so were dropped.

Type-II narrow emission-line galaxies, (NLAGN, class='N') are added as the luminosity class corresponding to the type-I AGN galaxies.  High-luminosity type-II NLQSO (class='K') correspond to the type-I quasars.  The NLQSO/NLAGN divider is the same luminosity/psf function which demarcates QSOs (optically core dominated) from AGNs (host dominated).  Type-II NLAGN include unquantified contamination by legacy NELGs/ELGs/LINERs and probably a few starbursts which eluded removal, so it serves as a catch-all category presented for completeness, rather than as a strict type-II class. 

Milliquas sub-classifies quasars by whether they are optically core dominated QSOs (class='Q') or host dominated AGN (class='A').  A new rule for processing SDSS quasars is that those with STARFORMING \& STARBURST subclasses, or GALAXY classified by the DR12Q superset, are now taken as being optically host dominated.  This is because those spectra show a strong stellar continuum base.  About 7000 objects are so reclassed.  
 
LAMOST DR5\footnote{http://dr5.lamost.org} pipeline quasars which are not included in the manual LAMOST DR5Q \cite{LAM5Q} are therefore anti-selected by that exclusion.  This is handled by processing them with the same rules as SDSS pipeline quasars, but the LAMOST subclass field is unpopulated, thus only those with radio/X-ray associations are accepted into Milliquas.  However, SDSS/WISE quasar candidates do show LAMOST pipeline redshifts as available.  
    
Where SDSS finds a different redshift for a LAMOST-discovered quasar, if the difference is $>$0.1z, the name \& discovery credit are transferred to SDSS.  There are only about 10 of those.  
 
A few gravitiationally lensed images are added as type='L', additional to the already-catalogued brightest one.  These are added only where the ASP \cite{ASP} optical data shows them, to account for any radio/X-ray associations to them.  

HMQ classified all QSOs/AGN of z$<$0.1 as AGN as an \textit{a priori} boundary to handle the absence of local quasars.  This boundary is now moved in to z=0.05 because of many core-dominated QSOs between z=0.05 and 0.1, e.g., IRAS 01267-2157 (SDSS J012910.99-214156.8) at z=0.93.  HMQ section 5 discusses these luminosity-based boundaries.

\section{QSO updates since the HMQ}

Positional fixes were done on $\approx$200 legacy objects, mostly moves of $<$5 arcsec.  Notably, the quasar "Q 1409+732" is moved to the co-ordinates stated by its discovery paper \cite{A1987} where a suitable object resides.  Confusion came from that paper's discrepant finding chart onto a plate flaw.  Also some other positional fixes are recounted in the ``Related Postings'' section below.

Quasar doublets are now presented with both objects in every case.  Sometimes the discovery paper only averred the 2nd quasar without giving any astrometry or photometry, thus necessitating multiple queries to obtain those.     

A new quasar was presented by Milliquas v6.2: SDSS J101012.77+560520.0, v=18.7, z=2.136.  This object eluded SDSS capture by being in a 2-arcsec doublet with a red star, SDSS J101012.65+560520.4.  SDSS-DR3\footnote{http://classic.sdss.org/dr7/products/spectra/getspectra.html} targetted the star and pipeline-classified it as a QSO (z=2.136) due to emission bleed-over from the true quasar, and it was so accepted by SDSS-DR3Q \cite{DR3Q} without noticing the doublet.  That star was later removed by SDSS-DR9Q \cite{DR9Q}, probably because of its stellar continuum, and wasn't revisited after.  I found this quasar whilst reconciling \textit{Gaia} astrometry to the doublet.  SDSS has been notified and they have advised that they may add it into their upcoming SDSS-DR16Q release.

\section{QSO candidate updates since the HMQ}
    
Additional QSO candidates were sourced from the NBCKDE v3 catalog \cite{NBCKv3} and a variability \& refraction sourced catalog \cite{PETERS}.  These added about 300K new candidates to Milliquas.
    
WISE quasar candidates were added from the AllWISE MIRAGN catalog \cite{SECREST}; they are ~430K candidates over the whole sky for which 2-colour optical objects were found within a 2-arcsec radius of the AllWISE detection.  I processed those into pQSOs from calibration against the SDSS-DR12Q multi-class superset, and obtained photometric redshifts using the four-colour based method from the HMQ appendix 2.  The four colours used were B-R, R-W1, W1-W2 \& W2-W3.  
    
A subset of quasar candidates are SDSS-listed eBOSS targets.  So far, SDSS-IV hasn't published those investigated targets which were found to be ``not quasars'', therefore there is a creeping anti-selection built into those candidates.  This is now handled by arbitrarily lowering the pQSOs of all eBOSS-targetted candidates (not otherwise qualifying) by squaring those pQSO ratios, e.g., 90\% $\rightarrow$ 81\%.  Consequently $\approx$57K eBOSS candidates have fallen below pQSO=80\% and so are dropped from Milliquas.  SDSS advises that DR16Q will publish a ``superset'' of all determined classifications in mid-year 2020, so this topic may be revisited then.

\section{QSOs dropped since the HMQ}

Milliquas excludes low-confidence/quality or questionable objects (so deemed by their researchers), but many such objects were inherited from VCV \cite{VCV} which was more forgiving.  They are removed as encountered but a residue remains.  The following list shows those recently removed.

\begin{itemize}
	\item 22 objects from \cite {BOYLE}, classified by them as "possible" QSOs with "uncertain" redshifts, were removed from Milliquas.
	\item 42 AGN were flagged by their discovery paper \cite{MAUCH} as having "not certain" classification, neither did any of those have stellar PSFs nor secure radio/X-ray associations, so they are dropped from Milliquas.  
	\item Reclassifications were done for Milliquas objects deemed by the SDSS-DR14 pipeline to be plain galaxies -- these were heavily spot-checked to confirm.  16 legacy QSOs, 172 AGN, and 2343 type-II objects were thusly found to be just galaxies and so were dropped, and 433 legacy QSOs were found to be host-dominated and so reclassified to AGN. 
	\item Blazar candidates with neither redshift nor radio/X-ray association, about 30 objects, were dropped.  Most were stated low confidence in legacy papers.    
	\item 223 objects from \cite{IOVINO} which had no quasar-like colour/PSF profile nor any radio/X-ray/WISE association were removed, leaving 917 in Milliquas.  This resolves a cautionary note given in the HMQ section 2.B.4 end.  
	\item A few duplicates were identified \& removed, notably SBS 1315+605 as duplicate to SDSS J131715.46+601533.1, with the original name being transferred over to the SDSS object.
\end{itemize}

\section{Radio/X-ray association updates since the HMQ} 

Radio/X-ray association likelihoods are now calculated at a granularity of 0.1 arcsecond astrometric offsets for \textit{Chandra}, \textit{XMM-Newton}, \& \textit{Swift} X-ray source catalogs and the FIRST radio source catalog.  

The 3XMM-DR8 and XMMSL2-2.0 Slew X-ray source catalogs\footnote{https://www.cosmos.esa.int/web/xmm-newton/xsa} were added and new X-ray associations calculated.  Also, high-confidence data from 2XMMi-DR3 were included, recognized as valid by 3XMM-DR5 \cite{ROSEN}, section 8.2.  

The Chandra ACIS source catalog \cite{WANG} was added and new X-ray associations calculated.  Also, the Chandra Source Catalog v2\footnote{http://cxc.harvard.edu/csc} is added, but only as a supplement to CSC v1.1 because v2 provides only stacked data for which the optical solution used by Milliquas cannot be calculated.        

The 2nd RASS source catalog 2RXS \cite{BOLLER} replaces 1RXS for the most part (excepting those analogous 1RXS sources which are securely associated to optical objects, with matching associations from other X-ray catalogues), with positional changes up to an arcminute being common.  The takeaway is that RASS isn't reliable for optical identification without additional evidence.
    
Double radio lobes indicate core activity of some kind, even if shielded from view.  These are calculated in Milliquas such that many are not identified elsewhere.  Thus I've elected to include all double-lobe associations of confidence$>$80\%, 3951 of which are onto classified galaxies included as type=G.

\section{Use of \textit{Gaia}-DR2 astrometry -- rules for inclusion.} 
  
\textit{Gaia}-DR2 \cite{GAIA2} astrometry is now used \& flagged for $\approx$63\% of Milliquas (MQ) objects; the ReadMe identifies the flag used.  The default astrometry from the 1.163G-object ASP catalogue \cite{ASP} can have large offsets even if rare, see its paper and Appendix A for a full discussion.  \textit{Gaia} sources were matched 1-to-1 with MQ objects on the critera that the \textit{Gaia} source is that nearest to the MQ object and that the MQ object is that nearest ASP optical to the \textit{Gaia} source.  Care was taken to avoid false matchings.  99\% of all matchings are within an arcsec offset, but to find valid farther matches I binned all matches by object class and offset distance in 0.1 arcsec bins, with hundreds of targetted spot checks done to refine offset limits and to check objects with anomalous \textit{Gaia} BP-RP colour suggestive of a false match.  SDSS quasar candidates could not be matched beyond 1 arcsec offset because they are often optically faint and/or in close groups for which \textit{Gaia} had a different object only.  Host-dominated AGN cores match to 1.5 arcsec beyond which \textit{Gaia}, oriented to point sources, often shows nearby stars only.  QSOs/Bl-Lacs match well out to within 4 arcsec; for those I spot-checked all matches beyond 2 arcsec offset, and all with off-colours beyond 1 arcsec offset, and identified \& removed 14 false matches.  However, for \textit{Gaia} sources without BP \& RP colours, QSOs match reliably only within 2 arcsec and Bl-Lacs within 1.5 arcsec offsets.  X-ray/radio-only candidates (unique to MQ) match reliably out to within 2.5 arcsec offset.  In all cases it was paramount to avoid false matchings, thus very many true matchings were lost beyond the offset cutoffs.  Table 1 bins the astrometric offsets of \textit{Gaia} data matched to the original ASP data used by Milliquas.

\begin{table}[t] 
\scriptsize	 
\caption{Counts of \textit{Gaia} sources matched to Milliquas objects, by astrometric offset -- ``0.1'' means 0.0$<$offset$\leq$0.1 arcsec, etc.}
\begin{tabular}{rrrr}
\hline 
\ offset & count &    & cum \\
\ (asec) & (MQv6.3) & pct & pct \\
\hline
0.1   &   802314  &   63.60  &   63.60  \\
0.2   &   266213  &   21.10  &   84.71  \\
0.3   &    80512  &    6.38  &   91.09  \\
0.4   &    43192  &    3.42  &   94.51  \\
0.5   &    24275  &    1.92  &   96.44  \\
0.6   &    15448  &    1.22  &   97.66  \\
0.7   &    10483  &    0.83  &   98.49  \\
0.8   &     7435  &    0.59  &   99.08  \\
0.9   &     5513  &    0.44  &   99.52  \\
1.0   &     3905  &    0.31  &   99.83  \\
1.1   &      575  &    0.05  &   99.88  \\
1.2   &      377  &    0.03  &   99.91  \\
1.3   &      242  &    0.02  &   99.93  \\
1.4   &      163  &    0.01  &   99.94  \\
1.5   &      138  &    0.01  &   99.95  \\
1.6   &      101  &    0.01  &   99.96  \\
1.7   &       73  &    0.01  &   99.96  \\
1.8   &       75  &    0.01  &   99.97  \\
1.9   &       71  &    0.01  &   99.97  \\
2.0   &       58  &    0.00  &   99.98  \\
2.1   &       50  &    0.00  &   99.98  \\
2.2   &       47  &    0.00  &   99.99  \\
2.3   &       29  &    0.00  &   99.99  \\
2.4   &       42  &    0.00  &   99.99  \\
2.5   &       24  &    0.00  &   99.99  \\
\hline
\end{tabular}
\end{table}

\section{Overview of Milliquas -- data and structure} 
     
Figure 1 shows the Milliquas sky coverage, dominated by the SDSS footprint.  The 2QZ stripe is seen at $\delta=-30^{\circ}$.  AllWISE candidates dominate other sky.        
     
\begin{figure} 
\includegraphics[scale=0.25, angle=0]{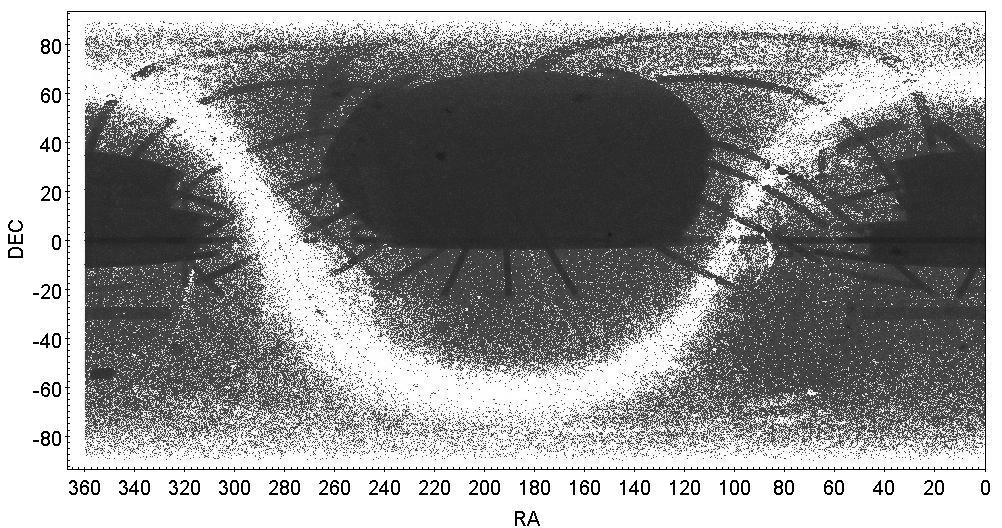} 
\caption{Sky coverage of Milliquas} 
\end{figure}

Table 2 shows the Milliquas data structure.  Each object is displayed with J2000 astrometry (usually from \textit{Gaia}-DR2 precessed to J2000 by CDS\footnote{https://cds.u-strasbg.fr/}), red and blue photometry, redshift, citations, and radio and X-ray associations where present.  The HMQ 4-digit citation numbers are retained and supplemented by 6-char citation identifiers which are indexed on the ReadMe.   

\begin{table*} 
\caption{Sample lines from Milliquas} 
\includegraphics[scale=0.325, angle=0]{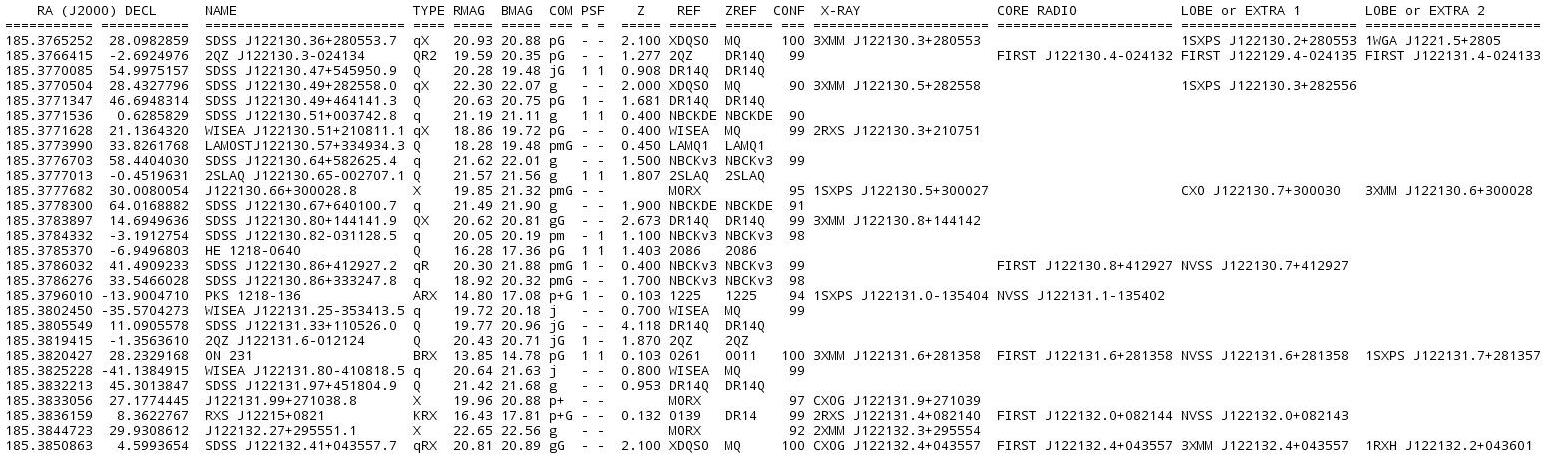} 
\tiny
Notes on columns (see ReadMe for full descriptions): 
\begin{itemize}
	\item TYPE: 1st char is the object classification: Q=QSO, A=AGN, B=BL Lac type, q=photometric, K=type II QSO, see ReadMe for full list.  Extra chars summarize the associations displayed: R=radio, X=X-ray, 2=double radio lobes. 
  \item COM:  comment on photometry: p=POSS-I magnitudes, so blue is POSS-I O, j=SERC Bj, g=SDSS green, +=optically variable, G=Gaia-DR2 astrometry, a=faint nuclear activity, m=nominal proper motion.
  \item PSF:  for red \& blue sources: '-'=stellar, 1=fuzzy, n=no psf available, x=not seen in this band.
  \item REF \& ZREF: citations for name and redshift; citations are indexed in the ReadMe \& "HMQ-references.txt" file.
  \item CONF: for candidates, calculated pct confidence that it is a quasar.  For classified objects, conf that the shown radio/X-ray detection is truly associated to it.  
  \item LOBE or EXTRA: if TYPE shows a '2' (=lobes), then double radio lobe identifiers are displayed here.  Otherwise, any additional radio and/or X-ray identifiers are displayed here.
\end{itemize}
The full table can be downloaded from http://quasars.org/milliquas.htm.
\end{table*}

Table 3 shows the top 20 contributing discovery papers ordered by numbers of name citations.  The large candidates catalogues show prominently. 

\begin{table}[t] 
\scriptsize	 
\caption{Top 20 discovery papers for Milliquas v6.4}
\tiny
\begin{tabular}{@{\hspace{0pt}}r@{\hspace{4pt}}l@{\hspace{2pt}}r@{\hspace{2pt}}r@{\hspace{4pt}}l}
\hline 
\  &    & \# of  & \# of \\
\# & ID & objects & redshifts & paper \\
\hline
 1 & SDSS DR14Q manual    & 513020 & 525745 & P\^{a}ris et al. \shortcite{DR14Q} \\ 
 2 & NBCKDE-v3 candidates & 468415 & 474810 & Richards et al. \shortcite{NBCKv3} \\
 3 & AllWISE candidates   & 432070 & 0      & Secrest et al. \shortcite{SECREST} \\
 4 & SDSS DR16 pipeline   & 167131 & 192299 & Ahumada et al. \shortcite{DR16} \\
 5 & XDQSO candidates     & 159424 & 0      & Bovy et al. \shortcite{XDQSO} \\
 6 & MORX candidates      & 52040  & 0      & Flesch, E. \shortcite{MORX} \\
 7 & NBCKDE candidates    & 34598  & 34487  & Richards et al. \shortcite{NBCKDE} \\
 8 & 2QZ/6QZ              & 27521  & 24160  & Croom et al. \shortcite{CROOM04} \\
 9 & PGC$^{\dagger}$      & 13689  & 17     & Paturel et al. \shortcite{PGC} \\
10 & SDSS DR14 pipeline   & 12087  & 14374  & Abolfathi et al. \shortcite{DR14} \\
11 & Peters candidates    & 10555  & 9468   & Peters et al. \shortcite{PETERS} \\
12 & 2SLAQ                & 10365  & 8709   & Croom et al. \shortcite{CROOM09} \\
13 & Milliquas            & 8684   & 537048 & data unique to Milliquas \\
14 & LAMOST QSO DR5       & 7998   & 7997   & Yao et al. \shortcite{LAM5Q} \\
15 & LAMOST QSO DR3       & 6771   & 6661   & Dong et al. \shortcite{LAM3Q} \\
16 & SDSS DR7Q manual     & 2089   & 329    & Schneider et al. \shortcite{DR7Q} \\
17 & AGES survey          & 2046   & 2046   & Kochanek et al. \shortcite{AGES} \\
18 & AAOz survey          & 1491   & 1498   & Lidman et al. \shortcite{AAOz} \\
19 & DEEP2 Redshifts      & 1435   & 1395   & Newman et al. \shortcite{NEWMAN} \\
20 & 3HSP blazars         &  807   & 1020   & Chang Y.-L. et al. \shortcite{3HSP} \\
\hline
\multicolumn{5}{l}{$^{\dagger}$ The Principal Galaxy Catalogue (PGC) is not actually a discovery} \\
   & \multicolumn{4}{l}{paper, but is used as a reference for names of AGN galaxies.} \\
\end{tabular}
\end{table}

\section{Related Postings on sci.astro.research}

Since publication of the HMQ (Flesch 2015), I sometimes posted on the newsgroup sci.astro.research on issues arising in the ongoing Milliquas releases, especially on further corrections of quasar positions.  In one of those, I described the recovery of the lost Cyril Hazard quasar "Q 0440-168" at (J2000) 04 42 40.30 -16 46 27.5, i.e., (B1950) 04 40 25.6 -16 52 05.  Extracts from a couple other postings are copied below to illustrate the kind of work done; they were written in an entertaining fashion to draw readers. \\\

10-January-2019 -- ``Tales of Cataloguing XIV -- the 0th finding chart'': \\

\small{\textsl{20th century quasar discovery papers made liberal use of finding charts to display the precise location of new quasars, lest the listed co-ordinates weren't accurate enough.  Astrometry had much improved by the 1990s but finding charts were still usually included, more as a tradition than a necessity.  That was all fine so long as the listed astrometry and finding chart agreed.  But sometimes they didn't.  Sometimes they pointed to different objects.}}  \\

\small{\textsl{I've given examples of this in earlier postings in this series, notably \#VIII "the log of jumping up \& down"  which is what you do when bad finding charts drive you crazy.  But other times it is good finding charts which save the day when the listed co-ordinates are false, e.g., the quasar "TOL 1038.2-27.1" from Bohuski \& Weedman 1979,ApJ 231,653, object \#23 (last one in the list), 41 arcsec offset from the false listed co-ordinates (unsuitable photometry r=18.0 b=20.1) to the true finding chart object (r=19.2 b=19.5).  Another example is the quasar "Q 0111-328" from Savage et al. 1984,MNRAS 207,393, which gave finding charts onto the original prism (grism) plates which are infallibly correct because the actual discovery spectrum is pointed at;  the B1950 co-ordinates given in the microfiche were of a nearby object offset by 76 arcsec.  Even big names like Schneider/Schmidt/Gunn did this for one object, "PC 0027+0515" in 1999, AJ 117,40, the Table 5 co-ordinates of which pointed near random objects whilst the true object was revealed on the finding chart at an offset of 17 arcsec.}}  \\

\small{\textsl{So there were bad finding charts and good finding charts.  But then there's this, from Borra et al. 1996, AJ 111,1456, the quasar "Q 13034+2942" (called "130324+294245" in the paper) is the very first one on their list, and their first finding chart.  They would make no error on the very first object, right?  Of course not.  Furthermore, on their Table 4a they lead right off with it as positioned at B1950 130324.21+294245.8, lest there be any mistake.  That translates to J2000 130547.40+292643.0 which shows up on the SDSS finding chart as a flattish-spectrum 21-mag stellar-psf which my own data reports as a variable object which was 19th magnitude in the 1960's -- so very quasar-like and all good.  So why did I previously have it catalogued as a reddish v=22 object 17 arcsec to the South-West?  Let's have another look at that finding chart, the first finding chart of the paper.}}  \\

\small{\textsl{There it is, but wait, they are pointing to the reddish object (which is probably a red dwarf star).  This is one of those finding charts where they don't use a photo, instead they re-create it with ink on paper.  Looks like they used a plotter (remember those?) and optical data.  Um, guys, your optical data did not include the true object.  It's not there at all.  They're pointing to the red dwarf because the quasar isn't on the chart.  Their very first finding chart for quasar discoveries points to a red dwarf star.  Looking at my archived catalogue versions, I originally had the right identification (inherited from VCV) but switched it to the red dwarf just before the publication of my Half-Million Quasars catalogue.  Guess I'd looked at one finding chart too many.}} ...  \\

\small{\textsl{On a separate note, the Soviet quasar "Q 0752+617" from Afanasiev/Lorenz/Nazarov 1989, SvAL 15,83 does not exist.  I've looked for it for years.  I've communicated with the lead author and he doesn't know where it is -- he knows only the old VCV location which is just the B1950 sky rectangle denoted by "0752+617".  There is no radio, no X-ray, no WISE candidate, no suitable bluish optical.  The paper stated narrow emission lines -- perhaps they measured a galaxy.  I give up, it is removed from the Milliquas catalogue as of the next edition.  I will gladly restore it if the authors provide its location.}}  (end) \\

Similarly, the unseen Soviet BAL quasar "SBS 1401+566" was dropped from Milliquas.  Its discovery paper 1986 Afanasiev/Erastova/Lipovetsky/Stepanian/Shapovalova, although cited as "in press" by Markarian/Stepanian/Erastova 1987-IAUS-121-25, was evidently never published.  This object was only an artifact of the literature.  \\\

29-January-2019 -- ``Tales of Cataloguing XV -- last fixes in from the cold'':  \\

\small{\textsl{(1) Abell 293 was originally thought to be a (Parkes) radio galaxy cluster but Gioia et al. 1984-ApJ-283-495, in addition to discovering a background X-ray quasar with z=1.897 and v=19.7, reported that "also present in this field is a strong point radio source PKS 0159+034 whose position is not coincident with any of the X-ray or optical sources discussed here".  In spite of this, VCV assigned both the name "PKS 0159+034" and the redshift 1.897 onto the primary Abell galaxy, thus conflating three objects into one.  Gioia et al. provided a finding chart of the quasar but didn't name it, however they placed it as the SE component of an Einstein-detected extended X-ray emission named "1E 0159.1+0330", therefore I've added this quasar (not previously catalogued) with name of "1E 0159.1+0330 SE".  Furthermore, my Milliquas algorithm shows the radio source PKS 0159+034, aka FIRST J020151.4+034309, to be associated with 98.6\% confidence to a r=21.8 g=22.1 stellar source which is furthermore calculated as 96\% likely to be a quasar.  But it has no redshift so will continue to appear in Milliquas as a candidate only, henceforth annotated with the name PKS 0159+034.}}  \\ 

\small{\textsl{(2) While checking over legacy quasars with unsuitable photometry, I came across the quasar "A4/22" from the "Very Faint Quasar Survey", D. Schade, 1991-AJ-102-869 -- the quasar is listed with z=1.045, v=20.00 and b=20.17.  But upon inspection on an SDSS finding chart, the near object is seen to be just a small passive galaxy, so how did David Schade come to call that a quasar?  I've seen this situation many times before, so I look for nearby objects, and I look North, South, East, West.  And sure enough, at 200 arcseconds due South of the designated spot I see SDSS J110205.85+295914.7 with g=20.00, r=19.71 and u=20.25, a perfect photometric fit.  Furthermore, it is a quasar candidate in Gordon Richards' NBCKDE-v3 catalogue (2015-ApJS-219-39) with a photometric redshift of 1.000, well-matched to Schade's spectroscopic redshift of 1.045.  Somehow this object got moved ~200 arcsec due North in the preparation of the paper -- looking at Table 4 of the paper, it looks like the declination of the object a4/5 was accidently copied over to a4/22.  These things happen.  I have moved the identification of "A4/22" over to J110205.85+295914.7, and the author has been informed.}} \\

\small{\textsl{(3) LMA 15, in the IC 1613 region, was surveyed by Lequeux/Meyssonnier/Azzopardi 1987-A\&AS-67-169 who provided a finding chart which however was too coarse to allow precise identification of the designated object.  The astrometry was given but only to whole time seconds, so had an uncertainty of $\approx$15 arcseconds.  VCV placed it at the given RA of B010239 (a faint star triplet), the true RA is found to be B010239.9, a blue stellar g=20.7, r=21.2 (PAN-STARRS) for a move of 13.7 arcseconds.}}  \\

\small{\textsl{(4) KP 1229.0+07.8 from Sramek \& Weedman, 1978-ApJ-221-468, \#19 on their list, z=1.93, v=20.5.  The B1950 astrometry, duly reported by VCV, pointed to nothing (approx 25 arcseconds away from 3 nearest optical candidates).  A coarse finding chart was provided from which, many years ago, I selected the wrong object, SDSS J123134.02+073440.2, a galaxy with r=19.25, g=20.61.  I have now moved this to the correct object, SDSS J123134.53+073425.8 with r=21.73, g=21.96, a move of 16.3 arcseconds.  Besides the flatter spectrum, it is also clearly a better match to the finding chart, although you do need to stare at it for a while.}}  (end)

\section{Conclusion}

The Milliquas (Million Quasars) catalogue v6.4 is presented as a complete record of published quasars to 11 December 2019, including the SDSS-DR16 pipeline release.  Milliquas presents 757\,991 type 1 QSOs \& AGN, $\approx$1.1M high-confidence (80\%+) photometric quasars, 2800 BL Lac objects, and 37\,116 type 2 objects.  Astrometry is 0.01 arcsecond accurate for most objects, and red-blue photometry is of 0.01 magnitude precision.  X-ray and radio associations for these objects are presented as applicable, including double radio lobes.

\begin{acknowledgements}
Thanks to non-commercial music (drone, etc.) for adding joy to the long hours.  This work was not funded.
\end{acknowledgements}

\end{document}